\documentclass[prd,aps,preprintnumbers]{revtex4}

\usepackage{graphicx}
\usepackage{bm}

\newcommand{\be}{\begin{equation}}
\newcommand{\ee}{\end{equation}}
\newcommand{\bea}{\begin{eqnarray}}
\newcommand{\eea}{\end{eqnarray}}

\newcommand{\calR}{{\mathcal R}}

\begin{document}

\preprint{YITP-08-80}

\setlength{\unitlength}{1mm}
 
\title{Effects of particle production during inflation}
\author{Antonio Enea Romano and Misao Sasaki}
\affiliation{Yukawa Institute for Theoretical Physics, Kyoto University,
Kyoto 606-8502, Japan}

\begin{abstract}
The impact of particle production during inflation on the primordial curvature
perturbation spectrum is investigated both analytically and numerically.
We obtain an oscillatory behavior on small scales, while on large scales
the spectrum is unaffected.
The amplitude of the oscillations is proportional to the number of coupled
fields, their mass, and the square of the coupling constant. 
The oscillations are due a discontinuity in the second time derivative of the
inflaton, arising from a temporary violation of the slow-roll conditions.
A similar effect on the power spectrum should be produced also
in other inflationary models where the slow-roll conditions are temporarily
violated.

\end{abstract}

\maketitle
\section{Introduction}
\label{Introduction}

Inflation is considered one of the most promising candidates to explain
the statistical features of the observable Universe revealed by the 
detection of the cosmic microwave background
(CMB)~\cite{Spergel:2003cb,Spergel:2006hy} and  large scale 
structure surveys such as the Sloan Digital Sky Survey
(SDSS)~ \cite{sdss}.

In the simplest models, inflation is an exponential expansion period of 
the Universe, driven by scalar field, called inflaton, slowly rolling 
down its potential.
One of its main predictions is a primordial curvature perturbation spectrum
of the form $P_R(k)=k^{n_s-1}$, where $n_s$ is the so-called scalar spectral
index. 

Various extensions of the simplest models have been proposed, and in this 
paper we will consider the coupling of the inflaton to another massive scalar
 field, which has been previously investigated by different groups, leading
 to apparently conflicting results~\cite{Chung:1999ve,Elgaroy:2003hp}.
The first group was in fact obtaining just a local peak in the power 
spectrum for scales which leave the horizon around the time of particle 
production, while the second obtained a small scale oscillatory behavior 
leading to a step between large and small scales.
We study the problem both analytically and numerically, obtaining an 
intermediate result which confirms the oscillation of the power spectrum
 on small scales, but without any step.
We also derive an analytical approximation for the curvature power spectrum,
 which clearly shows the dependency of the amplitude and period of the
 oscillations on the mass and number of the coupled fields and the 
coupling constant. 

We mention that the presence of a temporary non-slow-roll stage
during inflation and its effect on the scalar and tensor perturbation
spectrum as well as the resulting CMB anisotropy have been
investigated in \cite{Boyanovsky:2006pm,Destri:2008fj}. 
Our analysis may be regarded as a special case where the perturbation 
spectrum can be studied analytically, 
and confirms other general studies \cite{Starobinsky:1992ts,Gong:2005jr,Joy:2007na}
of the effects of singularities in the inflaton potential.
See also a recent paper by Joy et al.~\cite{Joy:2008qd} for comparizon with
the WMAP data.

The paper is organized as follows.
In Section~\ref{model}, we briefly describe the main features and motivations 
of the model we study.
In Section~\ref{calculation}, we describe our analysis and
present the numerical results for the spectrum of the curvature perturbation.
In Section~\ref{approximation}, we adopt an analytical approximation
and derive the spectrum in both large and small $k$ limits.
In Section~\ref{conclusion},
we summarize the results obtained and provide some ideas
 about possible extensions.
\section{model}
\label{model}

We will consider a theory with the potential,
\bea
V(\phi,\varphi) 
=V_0+\case{1}{2}m_\phi^2\phi^2+ \case{1}{2}N(m_{\varphi} - g\phi)^2\varphi^2\,,
\eea
where we assume $V_0$ dominates over the other terms during the period of
interest, so that the evolution of $H$ can be neglected in the calculation.

The equation for the inflaton can be written in the form \cite{Chung:1999ve}:
\bea
&&\ddot{\phi} +3H\dot{\phi} +m_\phi^2\phi -
            g N(m_{\varphi}-g\phi)\langle\varphi^2\rangle=0\,;
\\
&&\quad
\langle\varphi^2\rangle\approx\theta(t-t_0)
\frac{{\mathcal C}}{m_{\varphi}-g\phi}n_0\left( \frac{a}{a_0} \right)^{-3}\,, 
\quad
n_0=g^{3/2}\frac{|\dot{\phi}_0|^{3/2}}{(2\pi)^3}\,,
\label{varphisquare}
\eea
where ${\mathcal C}$ is a constant of order unity,
$t_0$ is the time at which $g \, \phi=m_{\varphi}$,
when the effective mass of the $\varphi$ field becomes zero and most of the 
particles are produced. Here and in the following, the suffix $0$ denotes
a quantity evaluated at $t=t_0$.
The two point function can be approximated with 
the expression above after renormalizing by subtraction of its asymptotic 
past value, when no particle were produced~\cite{Chung:1999ve}.

It can be shown that the time scale $\Delta t_c$ over which particle are 
produced is much smaller than the Hubble time, 
$H\Delta_c\sim 10^{-3}\ll 1$~\cite{Elgaroy:2003hp}. This justifies our 
approximation of $\langle\varphi^2\rangle$ by a step function as given by
Eq.~(\ref{varphisquare}) for scales $k/a_0<{\Delta t_c}^{-1}\sim 10^3H$.

For later convenience, we slightly rewrite the above field equation as
\begin{eqnarray}
&&\ddot{\phi} +3H\dot{\phi} +m_\phi^2\phi =S(t)\,;
\label{fieldeq}\\
&&\qquad
S=\theta(t-t_0)g N{\mathcal C}n_0\left( \frac{a}{a_0} \right)^{-3}\,.
\label{source}
\end{eqnarray}

We solve the equation for the curvature perturbation on co-moving
hypersurfaces:
\be
\label{eqr}
{\calR}_c''+2\frac{z'}{z}{\calR}_c+k^2 {\calR}_c=0\,;
\quad z\equiv \frac{a\dot\phi}{H}\,,
\ee
where a prime denotes the conformal time derivative, ${~}'=d/d\eta=a\,d/dt$.
The inflaton perturbation on flat hypersurfaces $\delta\phi_f$,
which is to be quantized on subhorizon scales, is related to $\calR_c$ as
\begin{eqnarray}
\delta\phi_f=-\frac{\dot\phi}{H}{\calR}_c\,.
\label{deltaphif}
\end{eqnarray}
We assume that there is no isocurvature perturbation after
inflation, So, the power spectrum of the curvature perturbation 
is given by
\be
\label{eqPr}
2\pi P_\calR^{1/2}(k)=\sqrt{2k^3}\,|\calR_k(t_f)|\,,
\ee
where $\calR_k$ is a properly normalized mode function,
and $t_f$ is the time at which inflation ends.

\section{calculation}
\label{calculation}

First we solve the inflaton background.
We assume the inflaton is slow-rolling before $t_0$.
Introduce two independent homogeneous solutions of (\ref{fieldeq}),
\begin{eqnarray}
U_{\pm}(t)=\exp[\lambda_{\pm}Ht]\,;
\quad \lambda_{\pm}=-\frac{3}{2}\left(1\mp\sqrt{1-4\mu^2/9}\right)\,,
\end{eqnarray}
where $\mu^2\equiv m_\phi^2/H^2$, and the Wronskian,
\begin{eqnarray}
W(t)\equiv\dot U_{+}\,U_{-}-\dot U_{-}\,U_{+}
=(\lambda_+-\lambda_-)HU_+U_-=(\lambda_+-\lambda_-)He^{-3Ht}\,,
\end{eqnarray}
the solution of our interest is expressed as
\begin{eqnarray}
\phi(t)=\phi_0\,U_+(t-t_0)+\int_{-\infty}^\infty dt'G(t-t')S(t')\,,
\label{phigensol}
\end{eqnarray}
where the Green function $G$ is given by
\begin{eqnarray}
G=\theta(t-t')\frac{U_+(t)U_-(t')-U_-(t)U_+(t')}{W(t')}\,.
\end{eqnarray}
Note that $U_+(t)$ describes the slow-roll solution, while
$U_-(t)$ the rapidly decaying solution; $\lambda_+\approx -\mu^2/3$
and $\lambda_-\approx-3+\mu^2/3$ for $\mu^2\ll1$.

With the source term given by Eq.~(\ref{source}), 
we obtain the solution explicitly as
\begin{eqnarray}
\phi(t)&=&\phi_0\, U_+(t-t_0)
\cr
&&+\frac{\theta(t-t_0)M^3}{\lambda_+\lambda_-(\lambda_+-\lambda_-)H^2}
\left[-\lambda_+U_+(t-t_0)+\lambda_-U_-(t-t_0)
+(\lambda_+-\lambda_-)e^{-3H(t-t_0)}\right]\,,
\label{phisol}
\end{eqnarray}
where we have introduced the mass scale $M$ by
\begin{eqnarray}
M^3\equiv gN{\mathcal C}n_0
=g^{5/2}N{\mathcal C}\frac{|\dot\phi_0|^{3/2}}{(2\pi)^3}. 
\end{eqnarray}

Taking the time derivative of this equation gives
\begin{eqnarray}
\dot\phi&=&\lambda_+H\phi_0\,u_+(t-t_0)
\cr
&&+\frac{\theta(t-t_0)M^3}{\lambda_+\lambda_-(\lambda_+-\lambda_-)H}
\left[-\lambda_+^2U_+(t-t_0)+\lambda_-^2U_-(t-t_0)
+(\lambda_+^2-\lambda_-^2)e^{-3H(t-t_0)}\right]\,.
\label{dotphisol}
\end{eqnarray}
We note that this result implies the presence of 
a step in $\dot\phi$. At late times, $H(t-t_0)\gg1$, the above
equation reduces to
\begin{eqnarray}
\dot\phi=\lambda_+H\phi_0\left(
1-\frac{M^3}{\lambda_-(\lambda_+-\lambda_-)H^2\phi_0}\right)U_+(t-t_0)\,.
\label{dotphilate}
\end{eqnarray}
Thus there is a step of the relative magnitude 
$\Delta\dot\phi/\dot\phi \sim M^3/(9H^2\phi_0)$ compared to the case
of no particle production, and it will be reflected in the overall 
shape of the spectrum in general.
However, as we shall see immediately below, for the values
of the parameters we choose, the step turns out to be negligible.

For ease of comparison, following~\cite{Chung:1999ve,Elgaroy:2003hp},
we will make the following choices for the model parameters:
\begin{eqnarray}
\label{eq:model1}
 m_\phi = 10^{-6}m_{pl}\,,\quad m_{\varphi}=2 m_{pl}\,,\quad g=1\,,
\quad V_0=5 m_\phi^2 m_\varphi^2  \,,
\end{eqnarray}
where $m_{pl}=G^{-1/2}$ is the Planck mass.
For this choice of the parameters, we find the step
relative to the case of no particle production is small,
$\Delta\dot\phi/\dot\phi \sim 10^{-5}N$ unless $N$ is
extremely large.
The behaviors of $\dot\phi$ and $\ddot\phi$
are plotted in Fig.~1 and 2, respectively.

\begin{center}
\begin{figure}[h]
\label{fig1}
\includegraphics{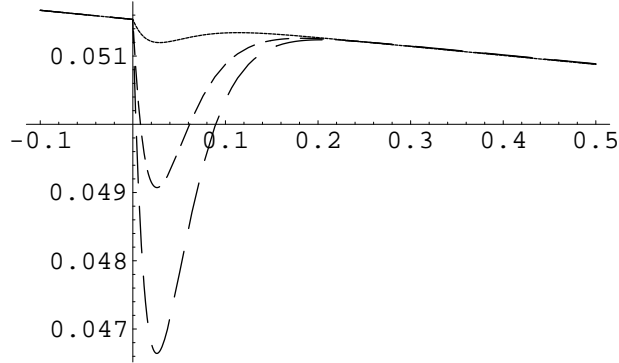}
\caption{$\dot\phi(t)/(m_p m_{\phi})$ is plotted for 
$-0.1<m_{\phi}(t-t_0)<0.5$.
The solid line corresponds to $N=1$,
the small dashed line to $N=8$, and the long dashed line to $N=16$. }
\end{figure}
\end{center}
\begin{center}
\begin{figure}[h]
\label{fig2}
\includegraphics{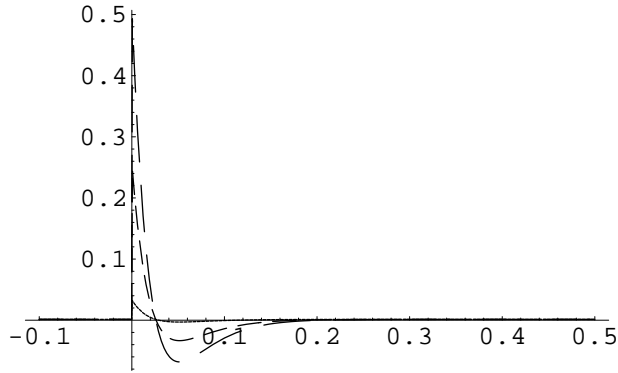}
\caption{$\ddot\phi(t)/(m_p m^2_{\phi})$ is plotted for 
$-0.1<m_{\phi}(t-t_0)<0.5$. The parameters are the same
as Fig.~1.}
\end{figure}
\end{center}

Setting $u\equiv a\,\delta\phi_f=-z\,\calR_c$, where
$z=a\,\dot\phi/H$, we have
\be
\label{uz}
 u''+\left(k^2-\frac{z''}{z}\right)u=0\,.
\ee
Since we assumed that the potential is dominated by $V_0$,
the time variation of $H$ in $z$ can be neglected and the scale factor
$a$ may be approximated by that of a pure de Sitter universe,
$a=(-H\eta)^{-1}$. On the other hand, the time variation of $\dot\phi$
cannot be neglected, particularly at and after the transition.
For $\dot\phi$, we use the solution given by Eq.~(\ref{dotphisol})
with the identification $Ht=\ln(-H\eta)$.

At $\eta=\eta_0$, $\ddot\phi$ is discontinuous.
Hence $z''$, which contains the third derivative $\dddot\phi$,
contains a delta function. This implies $u'$ is discontinuous
at $\eta=\eta_0$. To evaluate this discontinuity, we calculate
the contribution of the delta function in $z''$,
\begin{eqnarray}
D_0\equiv \int\limits_{\eta_0-\epsilon}^{\eta_0+\epsilon}{\frac{z''}{z}\,d\eta}
=\left[\ddot\phi_{0+}-\ddot\phi_{0-}\right]
\frac{a_0}{\dot{\phi_0}}=N\,g\,n_0\,{\mathcal C}\,\frac{a_0}{\dot{\phi}_0}
=\frac{M^3a_0}{\dot\phi_0}\,,
\label{D0}
\end{eqnarray}
where $\ddot\phi_{0-}$ and $\ddot\phi_{0+}$ are the values
of $\ddot\phi$ right before and after $t=t_0$, respectively.
Thus the matching condition at $\eta=\eta_0$ for $u$ is given by
\begin{eqnarray}
u'_{0+}=u'_{0-}(\eta_0)+D_0 u_{0-}\,,\quad u_{0+}=u_{0-}\,.
\label{match}
\end{eqnarray}
Turning back to the original variable $\calR_c$, it is noted
that this matching condition implies that 
both $\calR_c'$ and $\calR_c$ are continuous at $\eta=\eta_0$.
This is of course consistent with the evolution equation~(\ref{eqr})
for $\calR_c$, in which there is no delta function.

To calculate the power spectrum, we split it into two parts,
corresponding to the modes greater or smaller than $k_0$, where
$k_0=(aH)_0$. We assume the standard Bunch-Davies vacuum for
$\delta\phi_f$ at $\eta\to-\infty$, and solve for the positive 
frequency mode functions. Thus at sufficiently early times,
$\eta\to-\infty$, the mode function can be well approximated by
\begin{eqnarray}
\label{apprv}
u_{<}=v\equiv\frac{e^{-i k \eta}}{\sqrt{2k}}(1-\frac{i}{k\eta})\,,
\end{eqnarray}
where $u_{<}$ denotes the mode function at $\eta<\eta_0$.
For numerical analysis, we use this as the initial condition
for each mode when it is inside the horizon and when $\eta<\eta_0$.
Then we return to the original variable $\calR_c$ 
and numerically integrate Eq.~(\ref{eqr}). As we noted in the above,
there is no delta function in this equation, but only a discontinuity
in $z'/z$. Hence it can be numerically integrated across
the time $\eta=\eta_0$ without any problem.

For modes $k<k_0$, for which the sudden change in the effective
potential of the inflaton happens after horizon crossing,
Eq.~(\ref{apprv}) continues to be a good approximation until 
a mode crosses the horizon. Hence using it as the initial
condition at horizon crossing, we solve the differential equation
numerically. 

For modes $k>k_0$, the particle production takes place before the modes
leave the horizon. So, we set the initial condition at a sufficiently
early time $\eta=\eta_i<\eta_0$ common to all the modes, when the mode
functions are well approximated by Eq.~(\ref{apprv}). Then we
integrate Eq.~(\ref{eqr}) numerically.

\begin{center}
\begin{figure}[t]
\label{Neq8}
\includegraphics{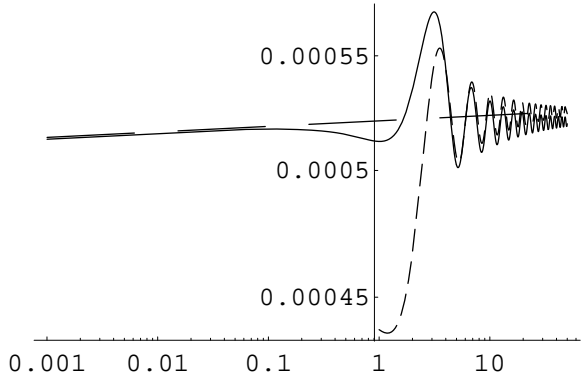}
\caption{ $P_{\calR}^{1/2}(k)$ is plotted for
 $3\times10^{-2}<k/(a_0 H_0)<50$ in the case of $N=8$. 
The solid line is the numerical result, the dashed line is 
the analytical approximation, and the long dashed line is the spectrum 
in the absence of particle production.}
\end{figure}
\end{center} 

\begin{center}
\begin{figure}[t]
\label{Neq16}
\includegraphics{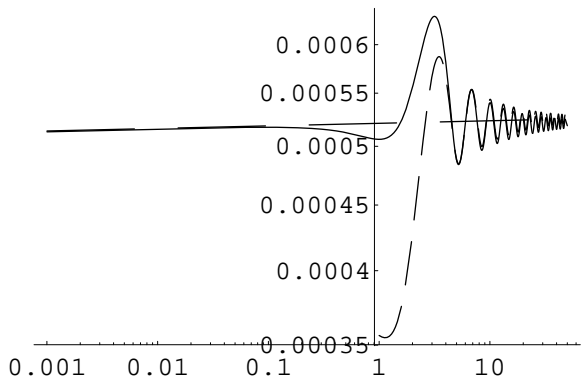}
\caption{The same as Fig.~3, but in the case of $N=16$.}
\end{figure}
\end{center}

The numerical results for the power spectrum are given in
 Figs.~3, 4 and 5.
For the modes $k<k_0$, we do not find any appreciable evolution 
of modes on super-horizon scales except for modes close to $k=k_0$.
For the modes $k>k_0$, the oscillatory feature of the spectrum is 
in agreement with \cite{Elgaroy:2003hp} and other 
studies such as \cite{Adams:2001vc}.

Our result that there appears no step in the spectrum is in agreement
with Adams et al.~\cite{Adams:2001vc}, while it differs from 
Elgaroy et al.~\cite{Elgaroy:2003hp} who claim that a step in the power
spectrum is produced due to the non-conservation of entropy perturbation.
We disagree with this interpretation, since a source term should be
present for the superhorizon evolution of $\calR_c$
due to entropy perturbations, which is absent in the present case.

In order to confirm our numerical results, we consider an analytical
approximation to the problem in the next section.
We find a good agreement between the numerical results 
and analytical approximations.

\begin{center}
\begin{figure}[t]
\label{Ncompare}
\includegraphics{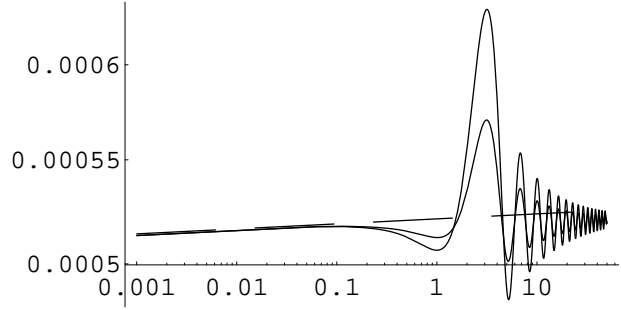}
\caption{$P^{1/2}_\calR(k)$ in the cases of both $N=8$ and $N=16$ 
are plotted for $3\times 10^{-2}<k/(a_0 H_0)<50$.
Clearly the amplitude of oscillations is larger for larger $N$,
approximately in proportion to $N$.
The long dashed line is the spectrum in the absence of coupling.}
\end{figure}
\end{center} 

\section{Analytical approximation of the spectrum}
\label{approximation}

In this section, we consider an analytical approximation for the
power spectrum. As in the case of numerical analysis, we split the 
modes into the two, $k<k_0$ and $k>k_0$, and discuss them separately.

For the modes $k<k_0$, since the transition occurs on superhorizon
scales, the only possible time variation of $\calR_c$ is 
due to the discontinuity in $z'/z$ in Eq.~(\ref{eqr}).
However, since $z'/z=a\,\dot z/z$, it grows exponentially
large as time goes on and the discontinuity becomes totally
irrelevant, unless the magnitude of the discontinuity
is exponentially large (which is apparently not the case).
Therefore, for $k\ll k_0$, there can be no time evolution of
 $\calR_c$ on superhorizon scales.
In fact, by using the technique developed in~\cite{Leach:2001zf},
one can explicitly show that the spectrum can be modified only near 
$k=k_0$ and quickly approaches the standard result at $k\ll k_0$.
Thus the curvature perturbation spectrum is given by
the standard formula,
\begin{eqnarray}
P_{\calR}^{1/2}(k)=\frac{H^2}{2\pi|\dot\phi(t_k)|}
\quad (k\ll k_0)\,,
\label{Psmallk}
\end{eqnarray}
where $t_k$ is the horizon crossing time, $k=a(t_k)H$.

As for the modes $k>k_0$, the analysis is a bit more complicated.
First we note that, a we stated before,
the mode functions $u$ are well approximated by
Eq.~(\ref{apprv}) when $\eta<\eta_0$. After the transition, 
there is a short period when the slow-rolling solution for the
background inflaton does not hold. Nevertheless, for sufficiently
large $k$, for which the mode is still deep inside the horizon
at the time of transition, this small violation of the slow-roll
conditions is totally negligible because $k^2\gg z''/z$.
Hence the approximate solution $v$ defined in Eq.~(\ref{apprv})
is still a valid solution even at $\eta>\eta_0$, including the
non-slow-roll period right after the transition.
However, $v$ is no longer the positive frequency mode function there.
Instead the desired mode function at $\eta>\eta_0$ should be
expressed as a linear combination of $v$ and $v^*$. 
Hence we may set
\begin{eqnarray}
u_{>}=\alpha_k\,v+\beta_k\,v^*\,,
\end{eqnarray}
where $u_{>}$ denotes the mode function at $\eta>\eta_0$.
It is useful to note that the coefficients $\alpha_k$ and $\beta_k$
can be expressed in terms of $u_{>}$, $v$ and $v*$ as
\begin{eqnarray}
\alpha_k=-i(v^*{}'u_{>}-v^*u_{>}')\,,
\quad 
\beta_k=i(v'u_{>}-vu_{>}')\,.
\label{alphabeta}
\end{eqnarray}
This gives the formula for the spectrum,
\begin{eqnarray}
P_{\calR}^{1/2}(k)=\frac{H^2}{2\pi|\dot\phi(t_k)|}|\alpha_k-\beta_k|\,.
\label{Pform}
\end{eqnarray}
It should be noted that $|\alpha_k|^2-|\beta_k|^2=1$.
This relation can be used as a consistency check.

Now, the matching condition~(\ref{match}) implies
\begin{eqnarray}
u_{>}(\eta_{0+})=v_0\,,\quad u_{>}'(\eta_{0+})=v_0'+D_0\,v_0\,.
\end{eqnarray}
Applying Eq.~(\ref{alphabeta}) to an epoch right after the 
transition, $\eta=\eta_{0+}$, with $u_{>}$ and $u_{>}'$
given by these equations, we find
\begin{eqnarray}
\alpha_k&=&1+iD_0v_0v_0^*
=1+i\frac{D_0}{2k}\left(1+\frac{1}{(k\eta_0)^2}\right)\,,
\cr
\beta_k&=&-iD_0v_0^2=-i\frac{D_0}{2k}
\left(1-\frac{i}{k\eta_0}\right)^2e^{-2ik\eta_0}\,.
\label{alphabeta}
\end{eqnarray}
Inserting them into Eq.~(\ref{Pform}), we obtain
the spectrum at $k\gg k_0$ as
\begin{eqnarray}
P_{\calR}(k)^{1/2}=\frac{H^2}{2\pi|\dot{\phi(t_k)}|}
\left(1+\frac{D_0}{k}\left[\sin2k\eta_0
+O\left(\frac{1}{k\eta_0}\right)\right]
+\frac{D_0^2}{2k^2}
\left[1+\cos2k\eta_0+O\left(\frac{1}{k\eta_0}\right)\right]\right)^{1/2}
\quad(k\gg k_0)\,.
\label{Plargek}
\end{eqnarray}
As seen from Figs.~3 and 4, the above analytical approximation
agrees well with the numerical results at large $k$.
However, the analytical approximation loses its accuracy
for modes which leave the horizon around the time of particle production 
$\eta_0$. This is because the effect of non-slow-rolling
is non-negligible for these modes and the approximate 
solution~(\ref{apprv}) is no longer valid.

The spectrum at $k>k_0$ behaves like a dumped harmonic oscillator,
where the amplitude of the oscillations is proportional to $D_0$,
provided $D_0/k<1$. As one can guess from the formula~(\ref{Pform}), together
with Eq.~(\ref{alphabeta}), these oscillations are due to an interference
between the positive and negative frequency mode functions, or 
contamination of negative frequency modes with positive frequency modes
due to the transition. One might doubt if this is due to the approximation
of replacing an otherwise smooth function by a step function
in Eq.~(\ref{varphisquare}). However, as we mentioned there, our approximation
is valid for $k< a_0/\Delta t=k_0(H\Delta t)^{-1}\sim 10^3 k_0$,
while the oscillations are present for all $k>k_0$. This confirms that
the oscillations are not an artifact of the approximation but real.

Using the expression for $D_0$, Eq.~(\ref{D0}),
the amplitude is evaluated as
\begin{eqnarray}
\frac{D_0}{k}=\frac{a_0}{k}\frac{M^3}{\dot\phi_0}
=\frac{a_0H}{k}\frac{Ng^2}{\sqrt{3}(2\pi)^3}
\frac{m_\phi\,m_\varphi^{1/2}}{H^{3/2}}\,.
\end{eqnarray}
This analytical estimate not only confirms the numerical result
that the oscillations amplitude is proportional to the number of
coupled fields $N$, but also shows that it is proportional to 
the square root of the mass 
$m_\varphi$ and is proportional to the square of the coupling constant $g$. 
It also shows that the period of the oscillations is given by $k_0\pi$.

\section{Conclusion}
\label{conclusion}

We have studied the impact of the coupling of the inflaton to
a scalar field on the primordial curvature perturbation spectrum.
We found the presence of an oscillatory behavior on small scales,
for the modes which leave the horizon after the time of particle production.
We have also presented a very good analytical approximation for the 
evolution of small scale modes. This method can be applied
to the general case of a sudden change in the inflaton potential,
leading to a temporary violation of the slow roll conditions.

The amplitude of the oscillations is proportional to the number of 
coupled fields $N$ and the square root of their mass $m_{\varphi}^{1/2}$, 
and to the square of the coupling constant $g^2$.
The period of the oscillations is $k_0\pi$, where $k_0$
is the wavelength that crosses the horizon right at the time of
the particle production.

On large scales, $k<k_0$, the power spectrum is
virtually unaffected by the particle production, contrary to what was 
claimed in a previous numerical investigation~\cite{Elgaroy:2003hp}.
This is because the violation of the slow-roll conditions
can affect only on those modes close to $k=k_0$ on superhorizon scales,
and the curvature perturbation is conserved on sufficiently large
superhorizon scales, $k\ll k_0$, no matter what occurs there.

Our results are quite general in the sense that
in models where slow-roll conditions are temporarily violated, 
the spectrum will have oscillations on scales smaller than 
the mode which leaves the horizon at time of transition, while
it will remain unchanged on the larger scales.
The presence of such features in the observed CMB spectrum could help to
 determine the magnitude and the lapse of periods during which
the slow-roll conditions are violated, although it may be difficult 
in practice to distinguish such a feature of a primordial origin
from a similar feature due to intermediate astrophysical processes 
happening before and/or after recombination. 

\begin{acknowledgments}
This work is supported in part by
Monbukagaku-sho Grant-in-Aid for the global COE program,
"The Next Generation of Physics, Spun from Universality and Emergence"
at Kyoto University.
EAR is supported by the JSPS postdoctoral fellowships.
MS is supported by JSPS Grant-in-Aid for Scientific 
Research (B) No.~17340075, and (A) No.~18204024, by JSPS 
Grant-in-Aid for Creative Scientific Research No.~19GS0219.
\end{acknowledgments}

\end{document}